\documentclass[onecolumn,showpacs,preprintnumbers,amsmath,amssymb,nofootinbib,APS]{revtex4}

\usepackage{dcolumn}
\usepackage{bm}
\usepackage[latin1]{inputenc}
\usepackage[spanish,english]{babel}
\usepackage{amsfonts}
\usepackage{amssymb}
\usepackage{graphicx}

\newcommand{\be}{\begin{equation}}
\newcommand{\ee}{\end{equation}}
\newcommand{\bea}{\begin{eqnarray}}
\newcommand{\eea}{\end{eqnarray}}

\begin{document}

\title{{\bf Two-point functions with an invariant Planck scale and thermal effects}}

\author{Iván Agulló and José Navarro-Salas}\email{ivan.agullo@uv.es  ,  jnavarro@ific.uv.es}
\affiliation{ {\footnotesize Departamento de Física Teórica and
IFIC, Centro Mixto Universidad de Valencia-CSIC.
    Facultad de Física, Universidad de Valencia,
        Burjassot-46100, Valencia, Spain. }}

\author{Gonzalo J. Olmo}\email{golmo@perimeterinstitute.ca}
\affiliation{ {\footnotesize Perimeter Institute for Theoretical
Physics, Waterloo, Ontario, N2L 2Y5 Canada}}

\author{Leonard Parker}\email{leonard@uwm.edu}
\affiliation{ {\footnotesize Physics Department, University of
Wisconsin-Milwaukee, P.O.Box 413, Milwaukee, WI 53201 USA}}

\date{April 2th, 2008}

\begin{abstract}
Nonlinear deformations of relativistic  symmetries at the Planck
scale are usually addressed in terms of modified dispersion
relations. We explore here an alternative route by directly
deforming the two-point functions of an underlying field theory. The
proposed deformations depend on a length parameter (Planck length)
and preserve the basic symmetries of the corresponding theory. We
also study the physical consequences implied by these modifications
at the Planck scale by analyzing the response function of  an
accelerated detector in Minkowski space, an inertial one in de
Sitter space, and also in a black hole spacetime.

\end{abstract}


\maketitle

\section{Introduction} One of the most profound principles of physics
is the principle of relativity, which establishes the physical
equivalence of all inertial observers. Discovered by Galileo in his
studies on the laws of motion, the principle of relativity was employed by Einstein, assuming as fundamental physical laws the
equations of electrodynamics, to propose the special theory of
relativity. The theory predicts a relation between the energy and
momentum of any body given by the well-known expression \be
\label{basic}  E^2 - p^2c^2 = m^2c^4 \ . \ee The theory was later
generalized in two directions. One was the incorporation of
the quantum uncertainty principle, which after many years gave rise
to the well established framework of quantum field theory. One of
the basic features of the relativistic quantum theory is the
existence of vacuum fluctuations, typically of the form  \be
\label{fieldcorrelations} \langle \Phi(x_1) \Phi(x_2) \rangle \sim
\frac{1}{4\pi^2}\frac{\hbar}{(x_1 - x_2)^{2}} \ , \ee as $x_1 \to
 x_2$.
 On the
other hand the principle of relativity was generalized by Einstein
to assume the physical equivalence of all freely falling observers,
when gravity is present, and culminated in the formulation of the
general theory of relativity. One of the main consequences of it is
the possibility of deforming the causal structure of  Minkowsky
spacetime. This happens, typically, in a gravitational collapse
producing a Schwarzschild black hole  \be ds^2 =
-(1-\frac{2GM}{c^2r})c^2dt^2 +
\frac{dr^2}{(1-\frac{2GM}{c^2r})}+r^2d\Omega^2 \ , \ee or, in a
cosmological  context, in  a de Sitter spacetime \be
\label{desittert}ds^2 = -c^2dt^2 + e^{2Ht}d\vec{x}^2 \ , \ee with a
cosmological constant $\Lambda = 3H^2/c^2$.

 The construction of a fully
consistent quantum gravity theory, incorporating both fundamental
theories in some limit, is one of the basic open problems of
theoretical physics. Despite this, there appears to be some robust
results when quantum field theory and general relativity
are combined at the semiclassical level \cite{birrel-davies, parker-toms}. The
thermal  properties associated with black hole and cosmological horizons \cite{hawking, gibbons-hawking}, with temperature \be
T=\frac{\kappa}{2\pi}\frac{\hbar}{ck_B} \ , \ee
where $\kappa$ is the surface gravity of the corresponding horizon, emerge as very deep
and robust results, and are likely to be fundamental to unravel the basic features of a quantum theory of gravity. The vacuum fluctuations (\ref{fieldcorrelations}) play a crucial role in the derivation of these thermal results.\\

A phenomenological way to explore the properties of a quantum
gravity theory, which is expected to incorporate the Planck scale
$l_P = (G\hbar /c^3)^{1/2}$,\footnote{For a general discussion and
earlier references, see \cite{garay}} is by modifying one of the
basic equations of special relativity, namely, the dispersion
relation (\ref{basic}), to incorporate  the Planck scale. At low
energies the dispersion relation is very well approximated by the
standard one, but when the energy or momentum saturates the Planck
scale the physics potentially departs from special relativity. There
are two different attitudes in doing this. One is to allow a
violation of the principle of relativity by distinguishing a
particular frame to which the modified dispersion relation refers.
This line of thought, motivated by condensed-matter analogies, was
originally aimed at explaining gamma ray phenomenology \cite{ellis}
and has led to the formulation of generally covariant theories of
gravity coupled
 to a dynamical timelike vector field \cite{jacobson}
(see, also \cite{mattingly}). A different attitude is to preserve
the principle of relativity, equivalence of all inertial frames, by
allowing a nonlinear action of Lorentz transformations keeping then
invariant the modified dispersion relation. This perspective was
pushed forward in \cite{amelino-camelia, magueijo-smolin} and has
some important connections to  previous mathematical studies
concerning quantum deformations of the Poincaré group
\cite{lukierski-nowicki-ruegg}.

In any case, although the modified dispersion relations are designed
to be applied to microscopic particles,  nothing prevents them from
being applied  to macroscopic bodies, as in conventional special
relativity. The Planck scale is then saturated immediately, giving
rise to some sort of paradox. Although this issue can be bypassed by
adding extra hypotheses, it strongly suggests that one should explore alternative
routes to deform special relativity. In this sense, one could
consider the deformation {\it ab initio} of an intrinsic quantum
mechanical object, with no classical analogue and, therefore, with
no possibility of being applied to conventional macroscopic
bodies. On the other hand, an important  lesson that follows from
working with deformed dispersion relations and non-linear
realizations of the Lorentz symmetry in momentum space,  is the
difficulty or perhaps impossibility of getting a proper realization of the kinematical symmetry in position space (independent of the energy-momentum
degrees of freedom). The non-linear action in the spacetime has been
implemented in the (eight-dimensional) extended phase-space. In
other words, one must double the number of dimensions
to find a non-linear realization of the Lorentz symmetry. We thus
find the following problem. How can one naturally consider quantum
objects, depending on eight variables, on which Lorentz
transformations may act non-linearly? The two-point correlation
functions of matter fields seem to be the simplest such quantum objects. The aim
of this paper is to explore this possibility. Firstly, as a
legitimate mathematical problem. But also by looking carefully at
its physical consequences concerning the thermal properties of
horizons.

In section \ref{sec:II}, we shall briefly review the main aspects of
the standard approach to introduce non-linear  actions of Lorentz
symmetry via deformed dispersion relations. This will help the reader to
better understand our proposal, in section \ref{sec:III}, to deform
the action of the kinematical symmetry via two-point correlation
functions. In this context, the natural thermal effect to be
analyzed with deformed actions of the Lorentz symmetry is the
acceleration radiation in Minkowski space. This will be done in
section \ref{sec:IV} by analyzing the response function of an
accelerated detector with a modified two-point function.
In sections \ref{sec:V} and  \ref{sec:VI} we shall extend our
considerations to de Sitter space and black holes, respectively.
Finally, in section \ref{sec:conclusions}, we shall briefly summarize
the main conclusions.

\section{ Deformed dispersion relations and nonlinear Lorentz actions \label{sec:II}}

Given a  modified dispersion relation of the general form (from now
on we shall assume $c=1$) \be E^2 f^2(E, p) - p^2 g^2(E, p)=m^2 \ ,
\ee the easiest way to modify the action of Lorentz transformations
on the energy-momentum $p^{\mu}$ is by introducing a nonlinear
invertible map $U$ between the physical quantities $(-E, p_i)$ and
an auxiliary energy-momentum $(-\epsilon, \Pi_i)$ \be U:\
p_{\mu}\equiv (-E, p_i) \ \to \Pi_{\mu}\equiv(-\epsilon, \Pi_i)=
(-f(E, p)E, \ g(E, p)p_i) \ . \ee While the auxiliary vector
$(-\epsilon, \Pi_i)$ transforms linearly under the Lorentz group,
the physical energy-momentum transforms as \be L( p^{\mu}) = [U^{-1}
\  L \ U] (p^{\mu}) \ . \ee The simplest choice for the function
$U^{-1}$ is \bea \label{U-1}E &=& \frac{\epsilon}{1 +
l_P\hbar^{-1}\epsilon }
\\ \nonumber p_i &=& \frac{\Pi_i}{1 + l_P\hbar^{-1} \epsilon} \ , \eea as was
proposed in \cite{magueijo-smolin} (it has a  deformed dispersion
relation of the form $(E^2 - p^2)/(1 + El_P\hbar^{-1})^{2}= m^2$).
In the limit $l_P \to 0$ the auxiliary energy-momentum coincides
with the physical one and the nonlinear action becomes the standard
linear transformation. Note also that, in special relativity the
energy $\epsilon$ is unbounded. Using arbitrarily large boosts one
can get $\epsilon \to +\infty$, but then the energy in the deformed
theory is saturated to the observer-independent Planck scale $E \to
\hbar/l_P$.

The above framework for nonlinear actions on momentum space can be
extended to the full (covariant) phase-space, parameterized by
$(p^{\mu}, x^{\mu})$, by further extending the map $U$ \be U: \
(p_{\mu}, x^{\nu}) \to  (\Pi_{\mu}, {X}^{\nu})  \ , \ee where \be
{X}^{\mu}= {X}^{\mu}(x^{\nu}, p_{\rho}) \ . \ee
 The above action ${X}^{\mu}(x^{\nu}, p_{\rho})$, and the corresponding one between physical variables $x'^{\mu}=(x^{\nu}, p_{\rho})$,  can be determined uniquely by imposing some extra physical condition. There are
essentially two different proposals in the literature. One demands
that plane-waves remain solutions of free field theories
\cite{kimberly-magueijo-medeiros}.
The other proposal requires that the full transformation be a
canonical one on phase-space \cite{mignemi, hinterleitner,
galan-mena-marugan}.

As remarked in the introduction, the proposal for realizing the bold
idea of deforming a relativistic theory in terms of deformed
dispersion relations has an apparent drawback. The deformation is
intended to be relevant for subatomic particles and in such a way
that either the energy and/or the momentum of the particle is
saturated at the Planck scale. But a formulation in phase space
applies equally to microscopic and macroscopic objects (or, in other
words, to quantum and classical objects). In the latter case it is
very easy to reach Planck energies, but Nature does not seem to
deform the established kinematical symmetries in this situation.
This fact motivates the approach of next section. Instead of working
initially in phase-space and deformed dispersion relations we shall
put forward an alternative way to further explore the physical
consequences of deforming the action of kinematic symmetries.
Instead of taking, as the starting point, a
 modification of dispersion relations we shall
consider,  ab initio,  a modification of the two-point function of a
free field theory, which makes sense only at the quantum level.

\section{Nonlinear actions and deformed two-point functions \label{sec:III}}

\subsection{Conformal field theories \label{sec:IIIA}}
To illustrate our approach it is  convenient to  consider the
largest kinematical symmetry allowed by a relativistic theory in a
generic $d$-dimensional flat spacetime. This is the conformal group
$SO(d,2)$. In a conformal field theory \cite{ginsparg} there is a
particular set of fields  with well-defined transformation laws
under conformal transformations $x \to x'$: \be
\label{Ctransformation} \Phi_j (x) \to \left|\frac{\partial
x'}{\partial x}\right|^{\Delta_j/d} \Phi_j (x') \ , \ee where
$|\frac{\partial x'}{\partial x}|$ stands for the jacobian of the
transformation and $\Delta_j$ is the dimension (or conformal weight)
of $\Phi_j(x)$. Restricting attention to the two-point correlation
function of a single field, covariance under the transformation
(\ref{Ctransformation}) implies\footnote{Unless explicitly stated
otherwise, all expectation values are computed with respect to the
vacuum state $|0\rangle$.}  \be \langle \Phi_j(x_1) \Phi_j(x_2)
\rangle = \left|\frac{\partial x'}{\partial
x}\right|^{\Delta_j/d}_{x_1}\left|\frac{\partial x'}{\partial
x}\right|^{\Delta_j/d}_{x_2}\langle \Phi_j(x'_1) \Phi_j(x'_2)
\rangle  \ . \ee As is well-known, invariance under Lorentz
transformations, translations, and dilations largely restricts the
form of the two-point function. One gets \be \langle \Phi_j(x_1)
\Phi_j(x_2) \rangle = \frac{C_j}{(x_1 - x_2)^{2\Delta_j}} \ , \ee
where $C_j$ is a constant related to the normalization of the field.
A typical example is a massless scalar field in $d=4$, for which
$\Delta = 1$ and then \be \langle \Phi(x_1) \Phi(x_2) \rangle =
\frac{1}{4\pi^2}\frac{\hbar}{(x_1 - x_2)^{2}} \ , \ee where  $ (x_1
- x_2)^2 \equiv -(T_2 - T_1)^2 + (X_2 - X_1)^2 + (Y_2 - Y_1)^2+ (Z_2
- Z_1)^2$.\footnote{When the two-point function is regarded as a
distribution one should replace, as usual, $T_2 - T_1$ by $(T_2 -
T_1 - i\epsilon )$} In $d=2$ the formalism should be refined since
then the global conformal group $SO(2,2)$ is enlarged to the
infinite-dimensional group of local transformation $x^{\pm} \to
x'^{\pm}(x^{\pm})$, where $x^{\pm}= t\pm x$ are null coordinates.
Relevant fields (usually called {\it primary fields}) should have a
weight $\Delta_+$ for $x^{+} \to x'^{+}(x^{+})$ and another one
$\Delta_-$ with respect to $x^{-} \to x'^{-}(x^{-})$. Typical
examples are the derivatives $
\partial_{\pm} \Phi $ of a two-dimensional massless scalar field
$\Phi$. In this case we also have $\Delta_+/d =1$ and $\Delta_-/d =
0$ for $\partial_+\Phi$, and the opposite weights for
$\partial_-\Phi$. This implies that \be
\label{2dcorrelations}\langle
\partial_{\pm}\Phi(x_1)
\partial_{\pm}\Phi(x_2) \rangle = -\frac{1}{4\pi}\frac{\hbar}{(x^{\pm}_1 -
x^{\pm}_2)^{2}}  \ . \ee

\subsection{Deforming the conformal two-point functions}

Let us now deform the action of the kinematical symmetry on
two-point functions mimicking the scheme followed above for modified
dispersion relations. We can introduce an invertible map $U$ defined
as \be U: \ \langle \phi_j(x_1) \phi_j(x_2) \rangle \ \to \langle
\Phi_j(x_1) \Phi_j(x_2) \rangle \ . \ee The action of conformal
transformations on the  correlations $\langle \Phi_j(x_1)
\Phi_j(x_2) \rangle$ induce, via $U$, an action on the (physical)
two-point functions $\langle \phi_j(x_1) \phi_j(x_2) \rangle $. This
can be seen with the following example. Let us choose the function
$U^{-1}$ similarly as in (\ref{U-1}) \be \label{U-12point}U^{-1}: \
\langle \phi_j(x_1)\phi_j(x_2) \rangle= \frac{\langle
\Phi_j(x_1)\Phi_j(x_2) \rangle}{1 - l_P^2 \hbar^{-1} \langle
\Phi_j(x_1)\Phi_j(x_2) \rangle} \ , \ee where the constant $l_P^2
\hbar^{-1}$, which (up to a factor $c^{-3}$) turns out to be equivalent to Newton's constant
$G$, is naturally chosen for dimensional reasons.  Accordingly,
the deformed action of conformal transformations is \be
\label{deformedaction} \langle \phi_j(x_1) \phi_j(x_2) \rangle \ \to
\ \frac{|\frac{\partial x'}{\partial
x}|^{\Delta_j/d}(x_1)|\frac{\partial x'}{\partial
x}|^{\Delta_j/d}(x_2)\langle \Phi_j(x'_1) \Phi_j(x'_2)
\rangle}{1-\l^2_P \hbar^{-1} |\frac{\partial x'}{\partial
x}|^{\Delta_j/d}(x_1)|\frac{\partial x'}{\partial
x}|^{\Delta_j/d}(x_2)\langle \Phi_j(x'_1) \Phi_j(x'_2) \rangle}\ .
\ee

\subsubsection{Case $d=4$}
For the massless scalar field in four dimensions the above formulas
leads to the following deformed two-point function \be \langle
\phi(x_1) \phi(x_2) \rangle =  \frac{\hbar}{4\pi^2(x_1 - x_2)^{2} -
l^2_P} \ , \ee and a modified action of conformal transformations
\be   \frac{1}{4\pi^2(x_1 - x_2)^{2} - l^2_P} \ \to \
\frac{|\frac{\partial x'}{\partial x}|^{1/4}(x_1)|\frac{\partial
x'}{\partial x}|^{1/4}(x_2)}{4\pi^2(x_1 - x_2)^{2} -
l^2_P|\frac{\partial x'}{\partial x}|^{1/4}(x_1)|\frac{\partial
x'}{\partial x}|^{1/4}(x_2)} \ . \ee Note that for the deformed
two-point function an invariant (Planck) scale emerges as\be \langle
\phi(x_1)\phi(x_2) \rangle|_{x_1 \to x_2} \approx
-\frac{\hbar}{l^2_P}= -\frac{1}{G} \ . \ee Furthermore, that
invariant quantity is not necessarily tied to the expectation value
in the vacuum state. If instead we have a different quantum state
$\Psi$, the Hadamard condition for the relativistic theory, namely
the universality of the short distance behavior \be \langle \Psi|
\Phi(x_1)\Phi(x_2)|\Psi \rangle|_{x_1 \to x_2} \sim \langle
\Phi(x_1)\Phi(x_2) \rangle|_{x_1 \to x_2} \ , \ee ensures  that \be
\langle \Psi |\phi(x_1)\phi(x_2)|\Psi \rangle|_{x_1 \to x_2} \approx
-\frac{\hbar}{l^2_P}= -\frac{1}{G} \ . \ee We would like to remark
that the invariant (observer-independent) scale $l_P^2\hbar^{-1}$
acts as a natural regulator for the two-point functions. This admits
a nice physical interpretation. When we probe quantum field theory
at scales $(x_1-x_2)^2\gg l_P^2$, gravity is negligible $(G\to 0)$
and the Green functions seem to diverge like $\sim1/(x_1-x_2)^2$.
For point separations of order $l_P$, the role of gravity in
constraining the space-time structure can no longer be neglected
$(G\neq 0)$, which results in $l_P^2\hbar^{-1}$ providing a natural
cutoff for the Green functions.

\subsubsection{Case $d=2$ \label{sec:IIIb2}}
The generic nonlinear realization (\ref{deformedaction}) can be
displayed more explicitly for the deformed correlations of the $d=2$
model mentioned above. The deformed correlations derived from
(\ref{2dcorrelations}) and (\ref{U-12point}) are then  \be \langle
\partial_{\pm}\phi(x_1)
\partial_{\pm}\phi(x_2) \rangle= \frac{-\hbar}{4\pi(x^{\pm}_1 - x^{\pm}_2)^{2} + l^2_P} \ . \ee
The deformed action of conformal transformations reads as \be
\langle
\partial_{\pm}\phi(x_1)
\partial_{\pm}\phi(x_2) \rangle \ \to \ \frac{\frac{dx'^{\pm}}{d
x^{\pm}}(x_1)\frac{dx'^{\pm}}{dx^{\pm}}(x_2)\langle \Phi_{\pm}(x'_1)
\Phi_{\pm}(x'_2) \rangle}{1-l^2_P \hbar^{-1} \frac{dx'^{\pm}}{d
x^{\pm}}(x_1)\frac{dx'^{\pm}}{dx^{\pm}}(x_2)\langle \Phi_{\pm}(x'_1)
\Phi_{\pm}(x'_2) \rangle}\ . \ee Therefore \be \frac{-\hbar}{4\pi
(x^{\pm}_1 - x^{\pm}_2)^{2} + l^2_P} \ \to \frac{-\hbar
\frac{dx'^{\pm}}{d x^{\pm}}(x_1)\frac{dx'^{\pm}}{dx^{\pm}}(x_2)}{
4\pi (x'^{\pm}_1 - x'^{\pm}_2)^{2} + l^2_P\frac{dx'^{\pm}}{d
x^{\pm}}(x_1)\frac{dx'^{\pm}}{dx^{\pm}}(x_2)} \ , \ee which clearly
shows the emergence of an invariant Planck scale. For instance,
under a boost of rapidity $\xi$: $x^{\pm} \to x'^{\pm}=e^{\pm
\xi}x^{\pm}$, the deformed two-point functions are \be
\frac{-\hbar}{4\pi (x^{\pm}_1 - x^{\pm}_2)^{2} + l^2_P} \ \to
\frac{-\hbar e^{\pm 2\xi}}{ 4\pi (x'^{\pm}_1 - x'^{\pm}_2)^{2} +
l^2_Pe^{\pm 2\xi}} \ , \ee so all inertial observers agree for the
coincident-point limit $-\frac{\hbar}{ l^2_P}$ of these correlation
functions.

\subsection{Massive scalar field}

The approach presented above for conformal field theories can be
extended easily to other theories. To illustrate this let us
consider a massive scalar field in four dimensions. One can deform
the corresponding two-point function in a way parallel to that used
for conformal theories. Taking into account that the two-point
function for a relativistic massive scalar field in four-dimensions
is \be \label{masive2p}\langle \Phi(x_1)\Phi(x_2) \rangle=\frac{m
\hbar }{4 \pi^2 \sqrt{(x_1 - x_2)^2}}K_1(m\sqrt{(x_1 - x_2)^2})\
,\ee where $K_1$ is a modified Bessel function, the deformed
two-point function reads
  \be
\label{deformation2point}\langle \phi(x_1)\phi(x_2) \rangle \equiv
\frac{\langle \Phi_m(x_1)\Phi_m(x_2) \rangle}{1 - l_P^2 \hbar^{-1}
\langle \Phi_m(x_1)\Phi_m(x_2) \rangle} = \frac{m\hbar
K_1(m\sqrt{(x_1 - x_2)^2})}{4\pi^2 \sqrt{(x_1 - x_2)^2} -m l_P^2
K_1(m\sqrt{(x_1 - x_2)^2})}\ . \ee  Note that,  since the
short-distance behavior of (\ref{masive2p})
 coincides with that of a massless field, the  invariant
(Planck) scale emerges in the same way \be  \langle
\phi(x_1)\phi(x_2) \rangle|_{x_1 \to x_2} \approx
-\frac{\hbar}{l^2_P} \ . \ee

\subsection{Relation to other approaches}

It is interesting to stress that a method to modify   the two-point
function in a Lorentz-invariant way has also been suggested in
\cite{padmanabhan1}, by invoking some sort of path-integral duality.
In terms of the Schwinger's proper time formalism, this approach is
equivalent to deforming the symmetric two-point
function\be \langle \Phi(x_1)\Phi(x_2) \rangle=
\int_{-\infty}^{+\infty} ds e^{-i m^2 s}K(x_1, x_2; s) = \frac{\hbar}{4\pi^2
(x_1 - x_2)^2 } \ , \ee where $K(x_1, x_2; s)$ is the heat kernel of
the matter field $\Phi$, as follows
   \be  \langle \phi(x_1)\phi(x_2) \rangle= \int_{-\infty}^{+\infty} ds e^{-i m^2 s}e^{il_P^2/(4\pi^2 s)}K(x_1, x_2; s)  \ , \ee where $K(x_1, x_2; s)$ is the same
relativistic heat kernel. For a massless field the above proposal
for deforming the  Green functions and that of (\ref{U-12point})
leads to the same deformed two-point function. However, one does not get the same result for generic massive fields.

\section{Response function of an accelerated detector: the role of
Planck scale \label{sec:IV}}

A natural way to show that the notion of particle is, in general,
observer-dependent was proposed in \cite{unruh1}. The particle
content of the vacuum perceived by an observer with trajectory
$x=x(\tau)$ and equipped with a detector, can be analyzed by
considering the interaction of the matter field $\Phi (x)$ with the
detector, modeled by the interaction lagrangian  (see, for instance,
\cite{birrel-davies}) \be g \int d\tau m(\tau)\Phi(x(\tau)) \ , \ee
where $m(\tau)$ represents the detector's monopole moment and $g$ is
the strength of the coupling. It is assumed that the detector has
some internal energy eigenstates $|E\rangle$, providing the internal
matrix elements  $\langle E |m(0)|E_0\rangle$ with the detector
ground state $|E_0\rangle$. For a general trajectory, the detector
will not remain in its ground state $|E_0\rangle$, but will undergo
a transition to an excited state $|E\rangle$. To first order in
perturbation theory, the transition amplitude from $|E_0\rangle
|0_M\rangle$ to $|E\rangle |\psi\rangle $, where  $|0_M\rangle$ is
the Minkowski
 vacuum state of the scalar field and $|\psi\rangle$ an arbitrary
field  state, is given by  \be g  \int d\tau \langle E \psi |m(\tau)\Phi(x(\tau))|
E_0 0_M\rangle \ . \ee The probability for  the detector to make the
transition from $|E_0\rangle$ to $|E\rangle$ (summing over all final
states of the scalar field)
 is then given by
\be g^2 |\langle E |m(0)|E_0\rangle|^2 F(E-E_0) \ , \ee
 where  $F(E-E_0)$ is the so-called response function of the
 detector
\be \label{FE}F(E-E_0)=
\int_{-\infty}^{+\infty}d\tau_1\int_{-\infty}^{+\infty}d\tau_2
e^{-i(E-E_0)(\tau_1 -\tau_2)/\hbar}\langle 0_M|
\Phi(x(\tau_1))\Phi(x(\tau_2)) |0_M\rangle \ .  \ee
When dealing with particular examples, one finds useful to introduce  the response rate function
\be \label{RRF}
{\dot F(E-E_0)} = \int_{-\infty}^{+\infty}d\Delta
\tau e^{-i(E-E_0)\Delta \tau/\hbar}\langle 0_M|
\Phi(x(\tau_1))\Phi(x(\tau_2))|0_M\rangle  \ . \ee
To work out the above expressions one should be careful when dealing with the typical
short-distance singularity of the two-point function. Usually one
considers the ``$i\epsilon$-prescription'' for the two-point function.\\

For a uniformly accelerated trajectory in Minkowski spacetime
\begin{equation} \label{accelerated}
t =\frac{1}{a} \sinh{a\tau} \ , \ x= \frac{1}{a} \cosh{a\tau} \ ,
\end{equation} where $a$ is the acceleration, the two-point
function, which for simplicity is chosen for a massless scalar
field, becomes \be \label{2pa}\langle 0_M|\Phi(x(\tau_1))\Phi(x(\tau_2))|0_M
\rangle= \frac{-\hbar
(\frac{a}{2})^2}{4\pi^2\sinh^2\frac{a}{2}(\Delta \tau -i\epsilon)} \
. \ee
Unlike for inertial trajectories, the response rate function ${\dot F(E-E_0)}$
of a uniformly accelerated observer does not vanish, and turns
out to be \be \label{rateF}{\dot F(E-E_0)}=
\int_{-\infty}^{+\infty}d\Delta \tau e^{-i(E-E_0)\Delta \tau/\hbar}
\frac{-\hbar(\frac{a}{2})^2}{4\pi^2\sinh^2\frac{a}{2}(\Delta \tau
-i\epsilon)} = \frac{1}{2\pi}\frac{E-E_0}{e^{2\pi (E-E_0)/\hbar a}
-1}\ . \ee This result tells us that a uniformly accelerated
observer in Minkowski space feels himself immersed in a thermal bath
at the temperature $k_B T = \frac{a \hbar}{2\pi}$. Note that even if
one considers the massive scalar, this result still holds. The mass
dependence of the two-point function does not affect the value of
the integral, which is only sensitive to the short-distance
behavior, already captured by the massless field.

We would like to remark that the use of the
``$i\epsilon-$prescription'' in treating the two-point function in
the distributional sense guarantees the vanishing of the response
function for an inertial observer. However, there is an alternative,
and more natural, way to enforce the vanishing of the response
function for inertial detectors. Essentially, we can calibrate the
detector with the accelerated (Rindler) vacuum $|0_R\rangle$ \be
\label{calibration} {\dot F(E-E_0)} =
\int_{-\infty}^{+\infty}d\Delta \tau e^{-i(E-E_0)\Delta
\tau}[\langle 0_M| \Phi(x(\tau_1))\Phi(x(\tau_2))|0_M\rangle -
\langle 0_R| \Phi(x(\tau_1))\Phi(x(\tau_2))| 0_R\rangle] \ . \ee The
integrand is now a smooth function, thanks to the universal
short-distance behavior of the two-point function, and the result is
equivalent to (\ref{rateF}) \be \label{rateFinertial}{\dot
F(E-E_0)}= \int_{-\infty}^{+\infty}d\Delta \tau e^{-i(E-E_0)\Delta
\tau/\hbar}\left[
\frac{-\hbar(\frac{a}{2})^2}{4\pi^2\sinh^2\frac{a}{2}\Delta \tau
}+\frac{\hbar}{4\pi^2(\Delta \tau)^2}\right]=
\frac{1}{2\pi}\frac{E-E_0}{e^{2\pi (E-E_0)/\hbar a} -1}\ . \ee
Although both expressions are mathematically equivalent, they lead
to different results when the correlation functions are deformed, as we will see later.

\subsection{Deforming the two-point function}

Let us now explore what happens when one deforms the standard
two-point function according to (\ref{deformation2point}). For
simplicity we shall considerer the case of a massless scalar field,
which leads to \be \label{deformation2pointmassless}\langle 0_M|
\phi(x_1)\phi(x_2)|0_M \rangle=  \frac{\hbar}{4\pi^2 (x_1 - x_2)^2 -
l_P^2} \ . \ee The rate ${\dot F(w)}$, where we define $w=E/\hbar$
and set $E_0=0$ for simplicity, can be worked out  in  a parallel
way to the standard relativistic theory. The  novelty is that one
has now  modified two-point functions. Therefore (\ref{RRF}) should
be replaced by (\ref{calibration}) and, accordingly,
(\ref{rateFinertial}) by \be \label{integralrateF} {\dot
F_{l_P}(w)}= -\frac{\hbar}{4\pi^2}\int_{-\infty}^{+\infty}d\Delta
\tau e^{-iw\Delta \tau}  \left[\frac{1}{(2/a)^2
\sinh^2(\frac{a}{2}\Delta \tau )+ l_P^2/4\pi^2} - \frac{1}{(\Delta
\tau)^2 + l_P^2/4\pi^2}\right]\ . \ee The final result is then \be
\label{rateFaf} {\dot
F_{l_P}(w)}=\frac{\hbar}{2\pi}\left[\frac{we^{\pi w/a}}{(e^{2\pi
w/a}-1)}\frac{\sinh[\frac{w}{a}(\theta-\pi)]}{\frac{w}{a}\sin\theta}
+\frac{\pi e^{-w l_P}}{l_P}\right]\ , \ee where $\theta\equiv
2\arcsin\left(\frac{l_P a}{4\pi}\right)$. The thermal Planckian
spectrum is smoothly recovered in the limit $l_P \to 0$. In fact,
the rate ${\dot F_{l_P}(w)}$ can be expanded as \be
\label{eq:therm-alpha} {\dot F_{l_P}(w)}\approx
\frac{\hbar}{2\pi}\left[\frac{w}{e^{2\pi w/a}-1} - \frac{l_P a^2
}{32\pi}+ O(l^3_P)\right] \ . \ee Thermality is maintained until
some frequency scale $\Omega$, which can be estimated by requiring
that the correction does not exceed the leading term in the above
approximated expression. A simple calculation leads to the condition
$32\pi\Omega e^{-2\pi \Omega/a} \approx l_P a^2$. Planck-scale
effects could potentially emerge at the scale $\Omega$, which is
roughly some orders above $T= a/2\pi $, if the acceleration is not
very high (in comparison with the Planck scale). This shows that the
thermal spectrum is essentially robust against trans-Planckian
physics. We would like to remark that the robustness of the effect
can also be explained within the standard relativistic field theory
framework \cite{ivan-navarro-salas-olmo-parker08} \footnote{In terms
of modified dispersion relations the robustness of the effect has
been addressed in \cite{rinaldi}}

Note that, if one ignores the calibration term, and naively works out
the response rate as
 \be
\label{naiveintegralrateF}
-\frac{1}{4\pi^2}\int_{-\infty}^{+\infty}d\Delta \tau e^{-iw\Delta
\tau}  \frac{\hbar }{(2/a)^2\sinh^2(\frac{a}{2}\Delta \tau )+
l_P^2/4\pi^2}\ , \ee the resulting expression \be \label{rate2}
\frac{w\hbar}{2\pi}\frac{e^{\pi w/a}}{(e^{2\pi
w/a}-1)}\frac{\sinh[\frac{w}{a}(\theta-\pi)]}{\frac{w}{a}\sin\theta}
\ , \ee largely departs from the thermal spectrum. Even worse,
 for an inertial observer, $a=0$, the response rate does not vanish, as one should expect
according to the principle of relativity. To produce a physically sound
result one must necessarily subtract the naive ``inertial'' contribution, as in
(\ref{rateFinertial}), replacing (\ref{naiveintegralrateF}) by
(\ref{integralrateF}).

\section{Thermal properties of de Sitter space \label{sec:V}}

It is well known that a geodesic observer in de Sitter space
(\ref{desittert}) feels a thermal bath of particles at a temperature
$k_B T = \frac{H \hbar}{2\pi}$, where $H=\sqrt{\frac{\Lambda}{3}}$.
One can derive this result \cite{gibbons-hawking} in a parallel way
to acceleration radiation effect of the previous section. We shall
now consider a scalar field in de Sitter space \be (\Box -
\frac{m^2}{\hbar^2} -\xi R)\Phi  =0 \ , \ee where $m$ is the mass
and $\xi$ stands for the coupling to the curvature. The response
rate function is also given by the general expression \be
\label{rateFGH}{\dot F(w)}= \int_{-\infty}^{+\infty}d\Delta \tau
e^{-iw\Delta \tau}\langle \Phi(x(\tau_1))\Phi(x(\tau_2)) \rangle \ ,
\ee where now the expectation value is understood with respect to
the (global) de Sitter vacuum $|0_{dS}\rangle$, invariant under the
$SO(4,1)$ isometries of the de Sitter spacetime. For simplicity,
without loss of generality, we shall restrict our discussion to
conformal coupling $\xi=1/6$ and $m=0$. In this situation the
two-point function takes the simple form \be \langle
\Phi(x_1)\Phi(x_2) \rangle=\frac{H^2\eta_1 \eta_2
\hbar}{4\pi^2[-(\eta_1 - \eta_2 -i\epsilon)^2+ (\vec{x}_1 -
\vec{x}_2)^2]} \ , \ee where $\eta= -H^{-1}e^{-Ht}$ is the conformal
time, for which the metric takes the form \be ds^2 =
\frac{1}{H^2\eta^2}(-d\eta^2 + d\vec{x}^2) \ . \ee Freely falling
detectors with trajectories \be t=\tau \ \ \ \ \ \vec{x}= \vec{x}_0
\ , \ee will have a response function with rate \be {\dot F(w)}=
\int_{-\infty}^{+\infty}d\Delta \tau e^{-iw\Delta \tau} \frac{-\hbar
H^2}{16\pi^2\sinh^2\frac{H}{2}(\Delta \tau -i\epsilon)}\ . \ee This
is exactly the same integral as (\ref{rateF}), producing now thermal
radiation at the temperature $k_B T = \frac{H \hbar}{2\pi}$.\\

\subsection{Deforming the two-point function}

Following the same strategy as for the uniformly accelerated
observer in Minkowski, we shall now deform the two-point function
preserving the de Sitter invariance $SO(4,1)$ instead of the
Poincaré invariance of the (global) Minkowskian vacuum. We can chose
the function $U$ as in  previous sections, which leads to \be
\label{U-12pointds}U^{-1}: \ \langle \phi(x_1)\phi(x_2) \rangle=
\frac{\langle \Phi(x_1)\Phi(x_2) \rangle}{1 - l_P^2 \hbar^{-1}
\langle \Phi(x_1)\Phi(x_2) \rangle} = \frac{H^2\eta_1 \eta_2
\hbar}{4\pi^2[-(\eta_1 - \eta_2 -i\epsilon)^2+ (\vec{x}_1 -
\vec{x}_2)^2] - l_P^2 H^2\eta_1 \eta_2} \ . \ee Armed with this
deformed two-point function we can evaluate the new response rate
function. As in  (\ref{calibration}), the integral for ${\dot
F_{l_P}(w)}$ involves the difference between the above correlation
function, evaluated in the global de Sitter vacuum, and the
two-point function evaluated in the vacuum of a comoving observer.
The two-point function of the comoving (inertial) observer $\langle
0_I|\Phi(x(\tau_1))\Phi(x(\tau_2))|0_I \rangle$ can be easily
obtained by summing over modes evaluated along the observer's
trajectory. To find the modes, one must solve the field equation \be
(\Box -\frac{1}{6}R)\Phi(\tilde{t},\tilde{\vec{x}})=0\ ,\ee where
$R=4 \Lambda=12 H^2$, using the static line element \be ds^2 = -(1-
H^2\tilde{r}^2)d\tilde{t}^2 + \frac{d\tilde{r}^2}{1-H^2\tilde{r}^2}
+ \tilde{r}^2 d\Omega^2 \ ,\ee where \be \tilde{t} =
-\frac{1}{2H}\ln [H^2(\eta^2 - r^2)] \ \ ,\ \ \ \tilde{r} = -
\frac{r}{H\eta} \ . \ee Along the observer's trajectory
$(\tilde{r}=0)$ only the s-wave modes\footnote{For completeness,
these modes are
$\Phi_{w,l=0}(\tilde{t},\tilde{r},\theta,\varphi)=\frac{1}{2 \pi
\sqrt{w}}\ e^{-i w \tilde{t}}\ \frac{1}{\tilde{r}}
\sin{\left[\frac{w}{H} \tanh^{-1}{(\tilde{r} H)}\right]}$}
contribute, and the two-point function becomes \be \langle
0_I|\Phi(x(\tau_1))\Phi(x(\tau_2))|0_I \rangle= \hbar \int_0^\infty
dw \frac{1}{4\pi^2 } w e^{-iw\Delta \tau}=-\frac{\hbar}{4 \pi^2
(\Delta \tau-i \epsilon)^2}\ . \ee The corresponding deformed
two-point function is \be \langle 0_I|\phi(x(\tau_1))\phi(x(\tau_2))
|0_I\rangle= -\frac{\hbar}{4 \pi^2(\Delta \tau)^2+l_P^2}\ . \ee It
is now immediate to see that the final result is equivalent to that
of the accelerated detector in Minkowski, equation (\ref{rateFaf}),
with the replacement $a \to H$.

\section{Black hole radiance and conformal symmetry\label{sec:VI}}

In this section we shall  stress that the Hawking effect can be
analyzed in the general framework presented in section II. Although
there are not global isometries in the spacetime of a collapsing
star, a powerful symmetry emerges in the near horizon region. This
region is effectively two-dimensional and conformally invariant, as
can be seen easily by writing the wave equation of a scalar field in
a Schwarzschild background. Expanding the field in spherical
harmonics \be \Phi(x^{\mu})= \sum_{l,m} \frac{\Phi_{l}(t,r)}{r}
Y_{lm}(\theta,\varphi) \ , \ee the four-dimensional Klein--Gordon
equation for $\Phi$ is then converted into a two-dimensional wave
equation, for each angular momentum component  \be
\label{equationf_l}\big ( -\frac{\partial^2}{\partial
t^2}+\frac{\partial^2}{\partial {r^*}^2} - V_{l}(r) \big
)\Phi_{l}(t,r)=0 \ , \ee with the potential
 $V_{l}(r)=\big (1-\frac{2GM}{r} \big ) \big
[l(l+1)/r^2+2GM/r^3 \big]$, and $r^*\equiv r + 2GM \ln |r/2GM -1|$
is the radial tortoise coordinate. In the near-horizon limit $r \to
2GM$ ($r^* \to -\infty$) the potential vanishes and
(\ref{equationf_l}) becomes the two-dimensional free wave equation,
 which in null  coordinates $(u\equiv t-r^*,v\equiv t+r^*)$ reads
$\partial_u \partial_v  \Phi_l =0$. This equation
 is very important since it exhibits the emergence of the
two-dimensional conformal invariance $u \rightarrow u'=u'(u), \ v
\rightarrow v'=v'(v)$, which is at the heart of the thermodynamic
properties of black holes. This symmetry plays a pivotal role in unraveling the
universal behavior of black hole entropy (see, for instance
\cite{carlip, skenderis, navarro}) and in the derivation of Hawking
radiation. To make explicit this latter fact we shall rewrite the
number of particles in terms of the two-point functions of a
two-dimensional conformal field theory. To be more precise, in the
conventional analysis in terms of Bogolubov coefficients, one first
performs the integration in distances (i.e., evaluation of the
scalar product between ``in'' and ``out'' modes) and leaves to the
end the sum of ``in'' modes. In contrast, we can invert the order
and perform first the sum of ``in'' modes, which naturally leads to
introduce the two-point function of the matter field, and leave the
integration in distances to the end \bea \label{Nijout} &&\langle
in|N_{i}^{out}|in\rangle =\sum_k \beta_{ik}\beta_{ik}^*=-\sum_k
(u^{out}_{i}, u^{in*}_k)(u^{out*}_{i}, u^{in}_k)=\nonumber \\
&=&\sum_k\left(\int_{\Sigma}d\Sigma^{\mu}_1u^{out}_{i}(x_1)
{\buildrel\leftrightarrow\over{\partial}}_\mu u^{in}_k(x_1)\right)
\left(\int_{\Sigma}d\Sigma^{\nu}_2u^{out
*}_{i}(x_2){\buildrel\leftrightarrow\over{\partial}}_\nu u^{in
*}_k(x_2)\right)\ , \ \eea where $\Sigma$ is an initial Cauchy
hypersurface. Taking into account that \be \langle in| \Phi
(x_1)\Phi (x_2)| in \rangle=\hbar\sum_k
u_k^{in}(x_1){u_k^{in}}^*(x_2) \ , \ee
we obtain a simple expression for the particle production number in
terms of the two-point function
\begin{equation}\label{eq:N-eps}
\langle in|N_{i}^{out}|in\rangle = \hbar^{-1} \int_\Sigma d\Sigma_1
^\mu d\Sigma_2 ^\nu
[u^{out}_{i}(x_1){\buildrel\leftrightarrow\over{\partial}}_\mu
][u^{out*}_{i}(x_2){\buildrel\leftrightarrow\over{\partial}}_\nu
]\langle in| \Phi (x_1)\Phi (x_2)|in\rangle \ ,
\end{equation}
where the ``$i\epsilon-$prescription'' is assumed to regulate the integrand. Alternatively, one can subtract $\langle
out|N_{i}^{out}|out\rangle \equiv 0$ from (\ref{Nijout}) to obtain
\begin{equation}\label{eq:N-ord}
\langle in|N_{i}^{out}|in\rangle = \hbar^{-1} \int_\Sigma d\Sigma_1
^\mu d\Sigma_2 ^\nu
[u^{out}_{i}(x_1){\buildrel\leftrightarrow\over{\partial}}_\mu
][u^{out*}_{i}(x_2){\buildrel\leftrightarrow\over{\partial}}_\nu
]\left[\langle in| \Phi (x_1)\Phi (x_2)|in\rangle-\langle out| \Phi (x_1)\Phi (x_2)|out\rangle\right] \ ,
\end{equation}
which is now regular provided $|in\rangle$ and $|out\rangle$ satisfy
the Hadamard condition\footnote{The two-point distribution should
have (for all physical states) a short-distance structure similar to
that of the ordinary vacuum state in Minkowski space:
$(2\pi)^{-2}(\sigma +2i\epsilon t + \epsilon^2)^{-1}$, where
$\sigma(x_1,x_2)$ is the squared geodesic
distance \cite{waldbook}.}.\\

 Let us apply this scheme in the formation process of a spherically
symmetric black hole (for details see
\cite{agullo-navarro-salas-olmo-parker}) and restrict the ``out''
region to $I^+$. The ``in'' region is, as usual, defined by $I^-$.
At $I^+$ the radial plane-wave modes are \be u^{out}_{wlm}(t,r,
\theta, \phi)|_{I^+}\sim u^{out}_{wl}(u)\frac{Y_{l}^m(\theta,
\phi)}{r} \ , \ee where $u^{out}_{wl}(u)= \frac{e^{-iwu}}{\sqrt{4\pi
w}} $. Integrating the angular degrees of freedom the emission rate,
evaluated at $I^+$, takes the form \bea \label{rateFbh}\langle
in|\dot{N}_{wlm}^{out}|in \rangle = \frac{|t_{l}(w)|^2}{\pi \hbar
w}\int_{-\infty}^{\infty}d(u_1-u_2) e^{-iw(u_1-u_2)} \langle in|
\partial_u \Phi_l(x_1)
\partial_u \Phi_l(x_2)|in \rangle \ , \eea
where  $t_l(w)$ represent the transmission coefficients  associated
with the potential barrier and \be \label{2pbh}\langle \partial_u
\Phi_l(x_1)
\partial_u \Phi_l(x_2) \rangle = -\frac{\hbar}{4\pi}\frac{\frac{dv}{du}(u_1)\frac{dv}{du}(u_2)}{[v(u_1)-v(u_2)-i\epsilon]^2}
 \ee is the two-point function of a two-dimensional conformal field
theory of the primary fields $\partial_u \Phi$ (see subsection
\ref{sec:IIIA}) transformed under the conformal
rescaling\footnote{Note that, due to the particular spacetime
geometry of a gravitational collapse, there is always a reflection
at $r=0$ which transforms $v \to u$. This explains why this
transformation is of the form $v \to v(u)$.} $u(v)\approx v_H-
\kappa^{-1}\ln \kappa (v_H-v)$, where $\kappa = 1/4GM$. Performing
the integration in $\Delta u \equiv u_1 - u_2$ we recover the
Planckian spectrum and the particle production rate \be \langle in
|\dot{N}^{out}_{wlm}|in \rangle= \frac{|t_l(w)|^2}{e^{2\pi
\kappa^{-1} w} - 1}\ . \ee

Now that Hawking radiation has been derived using two-point
functions, it is worth comparing equations (\ref{RRF}),
(\ref{rateFGH}), and (\ref{rateFbh}). In the three cases, the
two-point function is evaluated in the coordinates of the detector.
Let us see now how deformations of the two-point function would
affect Hawking radiation provided we use again the same deformation
function $U$. In this case, as explained in subsection
\ref{sec:IIIb2}, the effect of the Planck scale is encapsulated in the
deformed correlation function
\be \label{2pbhdef}
\langle \partial_u\phi_l(x_1)\partial_u \phi_l(x_2) \rangle = -\frac{\hbar}{4\pi}\frac{\frac{dv}{du}(u_1)\frac{dv}{du}(u_2)}
{[v(u_1)-v(u_2)]^2+l_P^2\frac{dv}{du}(u_1)\frac{dv}{du}(u_2)} \ ,
\ee
which preserves the conformal invariance by construction. Following
the analogy with the case of the  accelerated detector in Minkowski space
and the inertial one in de Sitter space, one should use expression (\ref{eq:N-ord})
to evaluate the emission rate in the modified theory. As the final result
we find again the function (\ref{rateFaf}), up
to the overall factor $w\hbar/2\pi$ and the grey-body coefficients,
with the replacement $a \to \kappa$

\be \label{ratebhaf}
\langle
in|\dot{N}_{wlm}^{out}|in\rangle =|t_l(w)|^2\left[\frac{e^{\pi
w/\kappa}}{(e^{2\pi
w/\kappa}-1)}\frac{\sinh[\frac{w}{\kappa}(\theta-\pi)]}{\frac{w}{\kappa}\sin\theta}
+\frac{\pi e^{-w l_P}}{wl_P}\right]\ .
\ee
where $\theta\equiv2\arcsin\left(\frac{l_P \kappa}{4\pi}\right)$.

\section{Conclusions \label{sec:conclusions}}

In this paper we have offered a unified view of the typical thermal
effects of quantum field theory in curved space aiming at showing
their robustness against trans-Planckian physics. We have shown that
all these effects can be easily derived in terms of two-point
correlation functions, which allowed us to explore the effects of
Planck scale physics by suitable deformations of such functions. The
deformations proposed here are somehow parallel the approach
presented in \cite{magueijo-smolin}, where dispersion relations were
modified while keeping the principle of relativity. In our case, the
two-point functions were deformed respecting the symmetries of the
original theory: Lorentz symmetry for the acceleration radiation
effect, de Sitter $SO(4,1)$ symmetry for the Gibbons-Hawking effect,
and the two-dimensional conformal symmetry for the Hawking effect.

One of the advantages of our approach is that it is relatively straightforward
to modify the theory maintaining the relevant symmetries of the problem and,
since our basic objects are two-point functions, the physical consequences can be
easily evaluated. On the other hand, it is worth noting that using the result
(\ref{rateFaf}), which corresponds to the acceleration radiation problem, we have been able to
show that the three effects described are robust under Planck-scale deformations.
This fact together with the elegant kinematic method of
connecting the acceleration, de Sitter and black hole radiation
given in \cite{deser-levin} may support our view that none of the
semiclassical thermal effects depends on ultra-high-energy physics.\\

This work has been partially supported by grants
FIS2005-05736-C03-03 and EU network MRTN-CT-2004-005104. L.Parker
has been partially supported by NSF grants PHY-0071044 and
PHY-0503366. I.Agulló and G.J.Olmo thank MEC for  FPU  and
postdoctoral grants, respectively.  J.Navarro-Salas thanks A. Fabbri
and M.Rinaldi for useful discussions. G.J.Olmo has been supported by
Perimeter Institute for Theoretical Physics. Research at Perimeter
Institute is supported by the Government of Canada through Industry
Canada and by the Province of Ontario through the Ministry of
Research \& Innovation.

\end{document}